\documentclass[fleqn,usenatbib,useAMS]{mnras}

\usepackage{graphicx}	
\usepackage{subcaption} 
\usepackage{xcolor}
\usepackage{soul}
\usepackage{amsmath}	
\usepackage{amssymb}	
\usepackage{multicol}        
\usepackage{bm}		
\usepackage{pdflscape}	
\usepackage[detect-all,per-mode=symbol]{siunitx}
\usepackage{ulem}

\DeclareSIUnit\year{yr}

\usepackage[T1]{fontenc}
\usepackage{ae,aecompl}

\usepackage{newtxtext,newtxmath}

\title{Is there a spectral turnover in the spin noise of millisecond pulsars?}

\author[B. Goncharov et al.]{
Boris Goncharov,$^{1,2}$\thanks{E-mail: \href{mailto:boris.goncharov@me.com}{boris.goncharov@me.com}}
Xing-Jiang Zhu$^{1,2}$
and Eric Thrane$^{1,2}$
\\
$^{1}$School of Physics and Astronomy, Monash University, Clayton, VIC 3800, Australia\\
$^{2}$OzGrav: The ARC Centre of Excellence for Gravitational Wave Discovery, Clayton, VIC 3800, Australia
}

\date{Last updated 2015 May 22; in original form 2013 September 5}

\pubyear{2019}

\begin{document}
\label{firstpage}
\pagerange{\pageref{firstpage}--\pageref{lastpage}}
\maketitle

\begin{abstract}
Pulsar timing arrays provide a unique means to detect nanohertz gravitational waves through long-term measurements of pulse arrival times from an ensemble of millisecond pulsars.
After years of observations, some timing array pulsars have been shown to be dominated by low-frequency red noise, including spin noise that might be associated with pulsar rotational irregularities.
The power spectral density of pulsar timing red noise is usually modeled with a power law or a power law with a turnover frequency below which the noise power spectrum plateaus.
If there is a turnover in the spin noise of millisecond pulsars, residing within the observation band of current and/or future pulsar timing measurements, it may be easier than projected to resolve the gravitational-wave background from supermassive binary black holes. Additionally, the spectral turnover can provide valuable insights on neutron star physics.
In the recent study by Melatos and Link, the authors provided a derivation of the model for power spectral density of spin noise from superfluid turbulence in the core of a neutron star, from first principles. The model features a spectral turnover, which depends on the dynamical response time of the superfluid and the steady-state angular velocity lag between the crust and the core of the star.
In this work, we search for a spectral turnover in spin noise using the first data release of the International Pulsar Timing Array.
Through Bayesian model selection, we find no evidence of a spectral turnover.
Our analysis also shows that data from pulsars J1939+2134, J1024$-$0719 and J1713+0747 prefers the power-law model to the superfluid turbulence model.

\end{abstract}




\begingroup
\let\clearpage\relax
\endgroup
\newpage

\section{Introduction}
\label{sec:intro}
It has long been proposed that pulsars can be used to detect gravitational waves in the nHz band \citep{sazhin1978gwb,detweiler1979pulsartiminggwb,hellingsdowns1983}.
Millisecond pulsars, first discovered in 1982 \citep{backer1982millisecond}, provide promising prospects for gravitational wave detection thanks to their exceptional rotational stability.
The concept of a pulsar timing array (PTA), long-term monitoring of pulse arrival times from a spatial array of millisecond pulsars, was conceived three decades ago \citep{Romani89,FosterBacker90}.
Currently, several collaborations are conducting PTA observations, including the Parkes Pulsar Timing Array (PPTA) \citep{manchester2013parkes}, the European Pulsar Timing Array (EPTA) \citep{kramer2013epta} and the North American Nanohertz Observatory for Gravitational Waves (NANOGrav) \citep{mclaughlin2013NANOGrav}. A consortium of these collaborations is called the International Pulsar Timing Array (IPTA) \citep{hobbs2010ipta,IPTAdr2}.

The first gravitational-wave signal detected with PTAs is likely to be a stochastic gravitational-wave background, formed by a cosmic population of supermassive binary black holes \citep{rosado2015firstdetectiongwb}.
Apart from the detection of gravitational waves, PTAs also offer the opportunity to establish a pulsar-based time standard \citep{hobbs2012pulsarclock}, to study the Solar System \citep{caballero2018studying}, the interstellar medium \citep{coles2015ismscattering} and the Solar wind \citep{madison2019solarwind}, and to constrain ultralight dark matter candidates \citep{porayko2018pptadm}.

The science output of PTA data relies on how well we model noise.
Incorrect noise models can also lead to false detection in gravitational-wave searches~\citep{11yrlimits2018nanograv,hazboun2019statistics}.
At low frequencies, where we are most sensitive to the stochastic gravitational-wave background, some millisecond pulsars, primarily studied in the PTA context, have measureable levels of red noise \citep{coles2011timingnoise, reardon2015timing, lentati2016spin, caballero2016noise, 9yeardata2015NANOGrav, 11yeardata2018NANOGrav}.
The red noise power spectrum is modelled by either a power law, or the broken power law, which introduces a corner frequency below which the noise power spectrum plateaus.
Additional opportunities also include the free spectral model \citep[see, e.g.,][]{lentati2013hyperefficient} and the power-law model with deviations at each frequency bin \citep{caballero2016noise}.
One particular source of red noise is the spin noise, which might be associated with pulsar rotational irregularities \citep[see, e.g.,][]{shannon2010spinnoise}.
While some young pulsars show hints of a spectral turnover at low frequencies \citep{parthasarathy2019timing}, it has not yet been found for millisecond pulsars.
If the typical time scale of a spectral turnover for millisecond pulsars is on the order of years or shorter, it reduces the red noise in the most sensitive frequency band of PTAs, yielding a faster detection of a stochastic gravitational-wave background.
Implications of how a spectral turnover will affect times to detection of a stochastic background were discussed in \cite{lasky2015pulsartiming}.
One of the conclusions of \cite{lasky2015pulsartiming} is that the gravitational wave power spectrum will only surpass the steeper timing noise spectrum if the latter flattens below some frequency.

Moreover, pulsar timing red noise provides interesting prospects for studying neutron star physics. A range of mechanisms have been proposed to explain pulsar red noise, including switching between two different spin-down rates \citep{lyne2010switchpdot}, recovery from a glitch -- a sudden increase in the rotational frequency \citep{johnston1999glitchrecovery}, a cumulative effect of frequent micro-glitches \citep{cordes1985microglitches, dAlessandro1995timingnoise,melatos2008microglitches}, variable coupling between the crust and liquid interior \citep{alpar1986spinnoisevortexcreep, jones1990spinnoisesuperfluid}, influence of planets \citep{cordes1993spinnoiseplanet} and asteroids \citep{shannon2013spinnoiseasteroid}. Nevertheless, there are not many models that link power spectral density model parameters to physical features.
One such model by \cite{melatos2013superfluid}, which we explore in this paper, predicts a superfluid turbulence in neutron star interiors as the origin of red noise. The turbulent process exerts a torque on the star's crust, where the external magnetic field of the star is produced. The model features a spectral turnover. 

In this work we employ Bayesian inference to search for evidence of spectral turnover in pulsar spin noise in the first data release (DR1) of the IPTA \citep{verbiest2016iptadr1}.
We discuss our data analysis methods in Section~\ref{sec:method}. Our simulation study is presented in Section~\ref{sec:simulations}. We describe the noise processes of the first IPTA data release in Section~\ref{sec:noiseipta}. We present the results in Section~\ref{sec:results}, and discuss our conclusions in Section~\ref{sec:conclusion}.

\section{Method}
\label{sec:method}

\subsection{Bayesian methodology in pulsar timing}
\label{sec:likelihood}

First, following~\cite{vanhaasteren2009ptabayesianmethod}, we assume a multivariate Gaussian likelihood function to describe pulsar timing residuals $\bm{\delta t}$ after fitting for the timing model:

\begin{equation}\label{eq:lnl1}
\begin{gathered}
\mathcal{L}(\bm{\delta t}| \bm{\theta},\bm{\xi}) = \frac{1}{\sqrt{(2 \pi)^n \text{det}(\bm{C})}} \\ \exp{\Bigg(-\frac{1}{2} (\bm{\delta t} - \bm{s} - \bm{M \xi})^{\text{T}} \bm{C}^{-1} (\bm{\delta t} - \bm{s} - \bm{M \xi})\Bigg)}~.
\end{gathered}
\end{equation}
Stochastic signals are modeled using a covariance matrix $\bm{C}$, while $\bm{s}$ is a deterministic signal vector. Parameters of our models are $\bm{\theta}$.
The vector $\bm{\xi}$ contains timing model parameters and $\bm{M}$ is a design matrix, describing the contribution of $m$ timing model parameters to $n$ times of arrivals (ToA).
Throughout our study, we work with ToAs and residuals, referenced to the Solar System Barycenter.
Assuming uniform prior on timing model parameters, the likelihood is marginalized over these parameters \citep{vanhaasteren2009ptabayesianmethod}:
\begin{equation}\label{eq:lnl2}
\begin{gathered}
\mathcal{L}(\bm{\delta t}|\bm{\theta}) = \frac{\sqrt{\text{det}(\bm{M}^{\text{T}} \bm{C}^{-1} \bm{M})^{-1}}}{\sqrt{(2 \pi)^{n-m} \text{det}(\bm{C})}} \\
\exp{\Bigg(-\frac{1}{2} (\bm{\delta t} - \bm{s})^{\text{T}} \bm{C}' (\bm{\delta t} - \bm{s})\Bigg)}~,
\end{gathered}
\end{equation}
where we have defined
\begin{equation}\label{eq:c1}
\bm{C}' = \bm{C}^{-1} - \bm{C}^{-1} \bm{M} (\bm{M}^T \bm{C}^{-1} \bm{M})^{-1} \bm{M}^T \bm{C}^{-1}~.
\end{equation}
To speed up the calculation, we employ the singular value decomposition of the design matrix in the form $\bm{M} = \bm{USV}^*$, where $\bm{S}$ contains singular values of $\bm{M}$, $\bm{U}$ and $\bm{V}$ are unitary matrices with dimensions $n \times n$ and $m \times m$ respectively. Then we obtain the likelihood function in a form \citep{vanhaasteren2012ptabayesianmethod2}
\begin{equation}\label{eq:lnl3}
\begin{gathered}
\mathcal{L}(\bm{\delta t}|\bm{\theta}) = \frac{1}{\sqrt{(2 \pi)^{n-m} \text{det}(\bm{G}^T \bm{C} \bm{G})}} \\ \exp{\Bigg(-\frac{1}{2} (\bm{\delta t} - \bm{s})^T \bm{G} (\bm{G}^T \bm{C} \bm{G})^{-1} \bm{G}^T (\bm{\delta t} - \bm{s})\Bigg)}~,
\end{gathered}
\end{equation}
so that $\bm{U} = \bm{U_1} \bm{G}$ with $\bm{U_1}$ and $\bm{G}$ consisting of the first $m$ and remaining $n-m$ columns of $\bm{U}$.

Some timing model processes are covariant with red noise. In particular, in analyses by \cite{coles2011timingnoise} and \cite{reardon2015timing}, the least-squares timing model fit absorbs some red noise. This absorption of power causes an apparent visible turnover in the measured spectra of red post-fit residuals, which is why the model with the broken power law was used for these analyses. In \cite{caballero2016noise}, the regular power-law was used, as the effects of timing model fitting were taken into account. In our analysis, we employ analytical marginalization over the uncertainty of timing model parameters in Equation~\ref{eq:lnl3}, which is equivalent to the simultaneous fitting of the timing model parameters and the red noise parameters, under the assumption that non-linear dependencies of the likelihood on the timing model parameters are negligible. This avoids the problem of detecting a spectral turnover that is actually due to the timing model fit, and makes it possible to target the spectral turnover in the spin noise itself. During marginalization, one loses sensitivity at low frequencies, especially at frequencies $ \leq 1/T_{\text{obs}}$, due to taking the uncertainty of the timing model into account.

Our prior probability distribution is $\pi(\bm{\theta})$. 
The integral of the likelihood times the prior over the prior parameter range is the Bayesian evidence for our model:
\begin{equation}\label{eq:evidence}
\mathcal{Z}(\bm{\theta}, \bm{\delta t}) = \int \mathcal{L}(\bm{\delta t}| \bm{\theta}) \pi(\bm{\theta}) d\bm{\theta}.
\end{equation}
To infer our model parameters $\bm{\theta}$, given observational data, we employ the Bayes' theorem:
\begin{equation}\label{eq:posterior}
\mathcal{P}(\bm{\theta}|\bm{\delta t}) = \frac{\mathcal{L}(\bm{\delta t}| \bm{\theta}) \pi(\bm{\theta})}{\mathcal{Z}(\bm{\theta}, \bm{\delta t})}.
\end{equation}
Using two different models A and B with parameters $\bm{\theta}_{\text{A}}$ and $\bm{\theta}_{\text{B}}$, we employ the Bayes factor as a measure of which model better fits the data:
\begin{equation}\label{eq:logbf}
\mathcal{B}_{\text{A},i}^{\text{B}} = \frac{\mathcal{Z}^{\text{B}}_i(\bm{\theta}_\text{B}, \bm{\delta t})}{\mathcal{Z}^{\text{A}}_i(\bm{\theta}_{\text{A}}, \bm{\delta t})},~i \in [1,N_\text{psr}]~,
\end{equation}
where $N_{\text{psr}}$ is the number of pulsars.
In Bayesian model selection, it is advised to use the posterior odds ratio as the decisive criterion for model comparison. Posterior odds ratio is equal to the Bayes factor times the prior odds ratio. In our model selection, we do not know a-priori whether the spectral turnover will ever be detected in millisecond pulsars. So, we choose prior odds to be equal to one. Thus, the posterior odds ratio is equal to the Bayes factor.
For simulation studies, we calculate the Bayes factors from evidence, which is obtained with nested sampling \citep{skilling2004nested}.
To save on computational cost, we adopt the product-space sampling method~\citep{hee2015bayesian,carlin1995bayesian} to calculate Bayes factors for the real data\footnote{The technical inconvenience of this method - one has to choose the set of compared models before the sampling starts - is the main reason to adopt nested sampling for our simulation studies.}.
Both methods are mathematically equivalent.
Assuming timing data for each pulsar are independent measurements, we combine all available data:
\begin{equation}\label{eq:logbftot}
\mathcal{B}^{\text{B}}_{\text{A}} = \prod_{i=1}^{N_{\text{psr}}} \mathcal{B}_{\text{A},i}^{\text{B}}\, ,
\end{equation}
which provides a metric to determine whether the spectral turnover is a real physical feature of millisecond pulsar spin noise.
For a discussion of how Bayes factors are combined through multiplication, see, for example, \cite{zimmerman2019combinelogbf}.
The authors argued that this approach is a limiting case of the inference of hyper-parameters that characterize the underlying distributions of parameters of individual events(pulsars), under the assumption that individual event (pulsar) parameters are independent.
We interpret Bayes Factors, as in~\cite{kass1995bayes}, where $0 \leq \log \mathcal{B} < 1$ is not worth more than a bare mention, $1 \leq \log \mathcal{B} < 3$ is positive, $3 \leq \log \mathcal{B} < 5$ is strong, and $\log \mathcal{B} \geq 5$ is very strong.

\subsection{Modelling stochastic processes}
\label{sec:stochastic_signals}
We model stochastic red noise processes as a power-law power spectral density $P(f)$.
We include $P(f)$ in our likelihood function using the Fourier-sum method from~\cite{lentati2013hyperefficient}, described briefly below.
We represent the covariance matrix as $\bm{C} = \bm{N} + \bm{K}$, where $\bm{N}$ is a diagonal matrix for white noise component, and $\bm{K}$ is a red noise component. A Woodbury lemma is used to simplify the inversion of a covariance matrix, decomposed into $\bm{N}$ and $\bm{K}$ \citep{hager1989woodburylemma,vanhaasteren2014newadvances}.
We define a Fourier basis $\bm{F}$ with elements:
\begin{equation}\label{eq:fourierbasis}
\begin{split}
F_{i,j}= 
\begin{cases}
\kappa_j~ a_i~\text{sin}\bigg(2 \pi f_i \Delta t_j \bigg), ~i ~\text{is even}~; \\
\kappa_j~ b_i~\text{cos}\bigg(2 \pi f_i \Delta t_j \bigg), ~i ~\text{is odd}~;
\end{cases}
\\ i \in [1,2 N_{\text{F}}], ~j \in [1,N_\text{ToA}]~.
\end{split}
\end{equation}
The parameter $\kappa$ is a constant, which we reserve to model chromatic red noise that depends on a radio frequency.
For spin noise, $\kappa$ is equal to one. 
The multiplicative factors $a_i$ and $b_i$ are Fourier coefficients which follow the standard Gaussian distribution. Each $\Delta t_j = (t_j-t_1)$ is the difference between the first ToA and the $j^{\text{th}}$ ToA.
The elements $f_i$ are components of a frequency vector that depend on the total observation span $T_{\text{obs}}$. They are defined as
\begin{equation}\label{eq:fouriercomponents}
f_i = 
\begin{cases}
\frac{i+1}{2T}, ~i ~\text{is odd}~; \\
\frac{i}{2T}, ~i ~\text{is even}~.
\end{cases}
\end{equation}
The variable $N_{\text{F}}$ determines the number of Fourier basis components in the frequency domain, with a minimum of $1/T_{\text{obs}}$ and spacing $\Delta f = 1/T_{\text{obs}}$.
Next, we obtain a diagonal matrix $\bm{\Phi}(\bm{\theta}_{\text{red}})$ with elements $\Phi_i = P(f_i)$, which depends on our red noise model with parameters $\bm{\theta}_{\text{red}}$.
Note, the minimum $f_i$ is sometimes referred to as the low-frequency cut-off, although it is not necessarily assumed that there is no red noise power below this frequency. Essentially, the data is just not analyzed below $f_i$. In principle, the low-frequency cut-off can become a free parameter of our model \citep[see, e.g.,][]{lentati2014temponest}. This approach could potentially reveal the sudden drop of power at low frequencies.
The red noise component in our likelihood function, marginalized over Fourier coefficients $a_i$ and $b_i$ \citep{vanhaasteren2014newadvances}, is
\begin{equation}\label{eq:getresiduals}
\bm{K} = {\bm{F \Phi  F}}^{\text{T}} ~\Delta f~.
\end{equation}
The white-noise covariance matrix $\bm{N}$ is diagonal with elements
\begin{equation}\label{eq:efacequad}
\sigma_j^2 = (\text{EFAC} ~\sigma^{\text{ToA}}_j)^2 + \text{EQUAD}^2~,
\end{equation}
where $\text{EFAC}$ and $\text{EQUAD}$ are factors to account for the excess of white noise, in addition to ToA error bars, $\sigma^{\text{ToA}}_j$.

\subsection{Red noise models}
\label{sec:red_noise_models}
Some millisecond pulsars in real data do not show evidence of red noise~\citep[e.g.,][]{lentati2016spin}.
We refer to the model without red noise as ``Model $\varnothing$''.
Next, we employ the two following phenomenological models for red noise. The power-law model
\begin{equation}\label{eq:default}
    P_{\text{PL}}(f) = \frac{A^2}{12\pi^2}\text{yr}^3(f ~\text{yr})^{-\gamma},
\end{equation}
which we refer to as the ``Model PL''. And the broken power-law model
\begin{equation}\label{eq:fancy}
    P_{\text{BPL}}(f) = \frac{A^2}{12\pi^2}\text{yr}^3(\sqrt{f^2+f_{\text{c}}^2} ~\text{yr})^{-\gamma},
\end{equation}
which we refer to as ``Model BPL''.
In the above two equations, model parameters are: the red noise amplitude $A$, the slope $\gamma$, the corner frequency $f_{\text{c}}$.

We also study the superfluid turbulence model from~\citep{melatos2013superfluid}
\begin{equation}\label{eq:melatos}
    P_{\text{M}}(f) = \frac{15 p^2}{8 \pi \lambda^2 \eta (R^{-1})} \int_{2 \pi}^{\infty} \frac{x^4 + 3 x^2 + 9}{[\frac{2 \pi f}{\eta (R^{-1})}]^2 + x^{4/3}} x^{-31/3} dx,
\end{equation}
which we refer to as ``Model M''.
The model depends on parameters $\eta(R^{-1})$ and $\lambda$. 
Our Equation~\ref{eq:melatos} is obtained by multiplying the power spectral density defined in  Equation 16 of \cite{melatos2013superfluid} with  pulsar spin period squared $p^2$.
This way, we obtain the power spectral density in units of $[s^3]$, to be consistent with Equations~\ref{eq:default} and \ref{eq:fancy}. 
Parameter $\lambda$ is a non-condensate fraction of the moment of inertia, which affects the amplitude of red noise. Parameter $\eta(R^{-1})$ is a decorrelation frequency, which determines the spectral turnover.
For convenience, we reparametrize Equation~\ref{eq:melatos}, in the form of parameters $M$ and $t_{\text{c}}$, using Equation~\ref{eq:reparam_melatos}.
The integral in Equation~\ref{eq:melatos} yields an analytical solution, given by Equation~\ref{eq:analytical_melatos}.
In our work, we do not model possible covariance between physical parameters $\lambda$ and $\eta(R^{-1})$, although they do implicitly depend on neutron star masses and radii, which are correlated~\citep{ozel2016masses}.
For this case, the more general approach from~\cite{zimmerman2019combinelogbf} for combining information from multiple measurements would be better suited.

\begin{figure}
\centering
\includegraphics[width=1.0\linewidth]{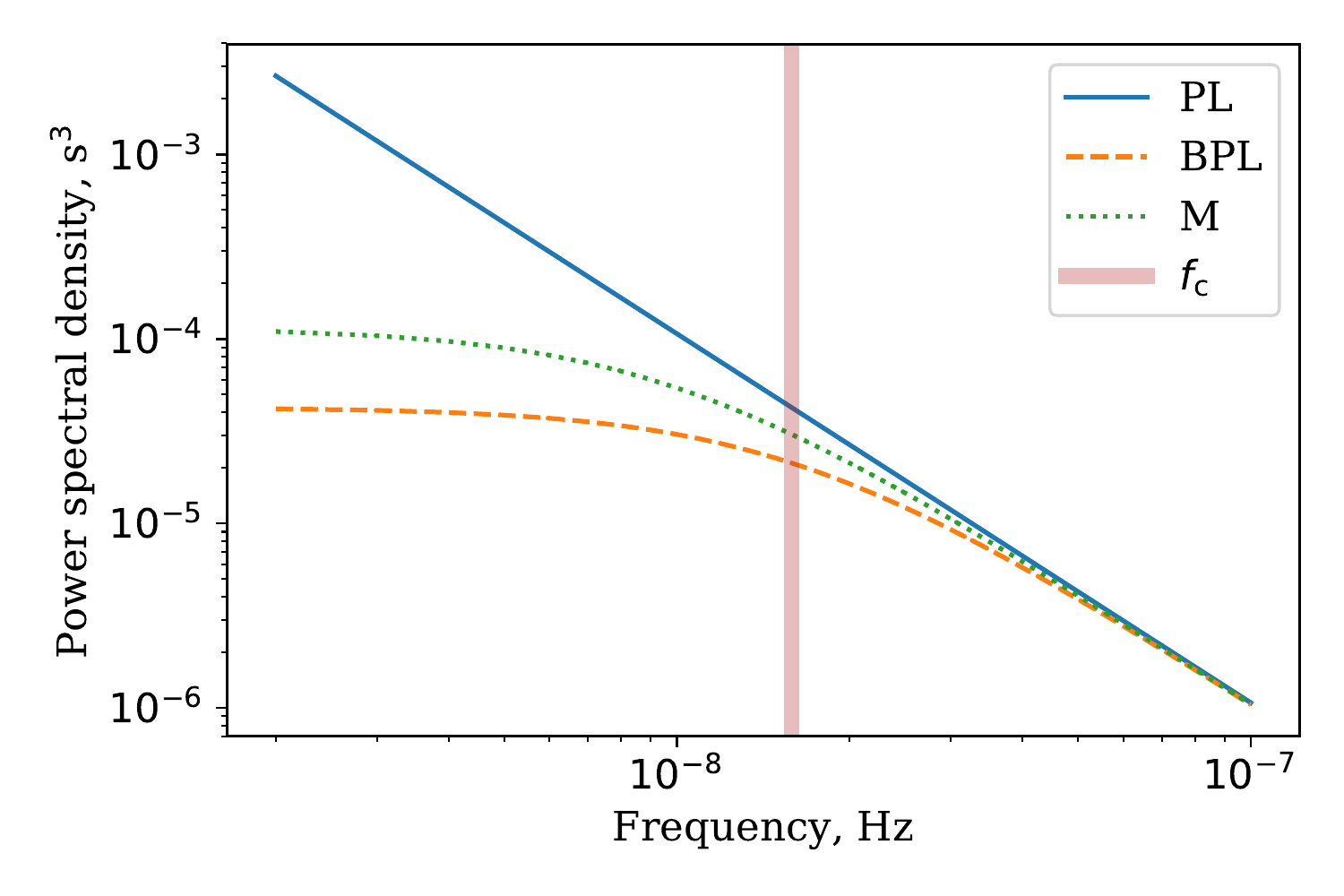}
\caption{
Models for pulsar red noise power spectral density. The blue solid line represents Model PL (Equation~\ref{eq:default}) and the orange dashed line represents Model BPL (Equation~\ref{eq:fancy}). For both of them we chose $A = 2 \times 10^{-13}$ and $\gamma = 2$. For the orange dashed line $f_{\text{c}} = {\SI{0.5}{\year^{-1}}}$. The green dotted line represents Model~\ref{eq:melatos} (Equation~\ref{eq:melatos}) with $\eta(R^{-1}) = {\SI{0.5}{\year^{-1}}}$, $\lambda = 0.5$, assuming pulsar spin period of ${\SI{1}{\milli \second}}$.}
\label{fig:spin_noise_models}
\end{figure}

In Figure~\ref{fig:spin_noise_models}, we plot examples of models of spin noise power spectral density.
Note, at high frequencies, Model M with two parameters asymptotically approaches Model PL with fixed $\gamma=2$ and only one free parameter (amplitude), so parameters $\eta(R^{-1})$ and $\lambda$ of Model M become degenerate.
In order to break this degeneracy, and to distinguish models PL and M, one must observe a spectral turnover.
This conclusion will be important later when we find pulsars that prefer Model M over Model PL, but realize that at the current stage of observations the performance of Model M is largely determined by the consistency of Model PL's estimate of $\gamma$ with 2.

In our analysis, we model $N_{\text{F}} = 30$ Fourier components of red noise processes. For power-law $P(f)$, the fraction of the signal power above $1/T_{\text{obs}}$ that is fit with $N_{\text{F}}$ components is equal to $1 - N_{\text{F}}^{1-\gamma}$ when $\gamma > 1$.
As an example, for a typical $\gamma = 3$, with 30 Fourier components  we take into account $99.9 \%$ of the red noise power above $1/T_{\text{obs}}$.
Below $\gamma = 1.5$, where 30 Fourier components take into account $81.7 \%$ of the red noise power above $1/T_{\text{obs}}$, it is better to use more Fourier components.
In reality, after we calculate this fraction for the power up to the sampling frequency, this fraction will be greater.
Nevertheless, for pulsar J2145$-$0750, where in \cite{lentati2016spin} it has been estimated that $\gamma = 0.6 \pm 0.2$, we use 100 Fourier components (107 components were used in \cite{lentati2016spin}).
We model remaining pulsars with 30 Fourier components, which is a reasonable and computationally-cheap approximation.
More comments on the consequences of this choice are provided in Section~\ref{sec:results}.

\subsection{Software}
\label{sec:software}
We estimate the design matrix using the \texttt{designmatrix} plugin in \texttt{TEMPO2} \citep{hobbs2006tempo2}. We simulate data and access \texttt{TEMPO2} using \texttt{libstempo} \citep{libstempo}. We construct our models and likelihood, and do parameter estimation using \texttt{Enterprise} \citep{enterprise2017}. We perform likelihood sampling using the \texttt{PTMCMCSampler} \citep{justin_ellis_2017_1037579} for IPTA DR1 data. For simulations we use a nested sampler \texttt{Dynesty} \citep{speagle2018dynesty}, and we use \texttt{Bilby} \citep{ashton2019bilby} to access the Dynesty sampler.

\section{Simulation study}
\label{sec:simulations}
We perform a simulation study to demonstrate our ability to do Bayesian model selection. We also demonstrate some potential subtleties in recovering a low-frequency turnover.
We simulate ToAs, ToA errors, and timing residuals for the pulsar J0711$-$6830, using ephemerides from the ATNF Pulsar Catalogue \citep{manchester2005psrcat}.
We simulate ToAs evenly sampled once every 30 days between MJD 53000 and 56650, which is roughly consistent with the average cadence of a typical IPTA observatory \citep[see][Table 1]{verbiest2016iptadr1}.
We assume ToA errors to be 0.5 $\mu$s, which is within the range of ToA errors as found in the first data release of the IPTA.
These parameters are applied to all simulations described in this section of the paper.
In our noise simulations we only assume one observing system, one observed radio frequency, and only red and white noise.
The red noise parameters chosen for simulations are described in the following subsections.
We choose them, so that they are approximately consistent with noise parameters of the real data \citep[see, e.g.,][Table 6]{lentati2016spin}.
The parameter values recovered from simulations in this section have been confirmed to be consistent with injected values.

\subsection{Red noise in an ensemble of pulsars}
\label{sec:sim_correct}
We simulate 50 mock pulsars with different random realisations of Model PL red noise and white noise. Then we perform model selection between Model PL and Model BPL. The simulated white noise parameters throughout the subsection are EFAC = 1 and EQUAD = 0.1 $\mu s$.
According to Section 3.3 of~\cite{verbiest2016iptadr1}, these are the typical EFAC and EQUAD values found in IPTA DR1.
The simulated red noise amplitude is different for the three cases we describe in this subsection, while the priors for red noise power-law index and corner frequency are $\pi(\gamma) = \mathcal{U}(2,5)$ and $\pi(f_{\text{c}}) = \text{log}_{10}~\mathcal{U}(10^{-10},10^{-6})$.
Here $\mathcal{U}$ stands for a uniform distribution, and $\text{log}_{10}~\mathcal{U}$ stands for a uniform in $\log_{10}$ distribution.
We use the same red noise priors for $A$ and $\gamma$ for models PL and BPL, for both injection and recovery.

First, we simulate Model PL with a prior $\pi(A) = \log_{10}~\mathcal{U}(10^{-14},10^{-11})$. The prior range for noise amplitude is chosen such that red noise is overall stronger than white noise. As a result, with all simulated pulsars, we obtain $\log\mathcal{B}^{\text{BPL}}_{\text{PL}} = -30.8$. Hence, Model PL is correctly preferred over Model BPL.

Second, we demonstrate that we do not prefer the wrong model if the red noise is overall much weaker than white noise. The prior for simulation and recovery of red noise amplitude is reduced to $\pi(A) = \log_{10}~\mathcal{U}(10^{-17},10^{-14})$. Now, $\log\mathcal{B}^{\text{BPL}}_{\text{PL}} = 1.0$. Therefore, if the red noise is too weak, we cannot distinguish between two models, as expected.

Finally, we demonstrate that, when the data from multiple pulsars are injected with Model BPL, our algorithm prefers Model BPL over Model PL.
To do this, we use the following prior on red noise amplitude $\pi(A) = \log_{10}~\mathcal{U}(10^{-14},10^{-11})$. Now we obtain $\log\mathcal{B}^{\text{BPL}}_{\text{PL}} = 96$ favouring the correct model.
Our results for this subsection are summarized in Table~\ref{tab:sim_correct}.
All injected signals were successfully recovered.

\begin{figure}
\centering
\includegraphics[width=1.0\linewidth]{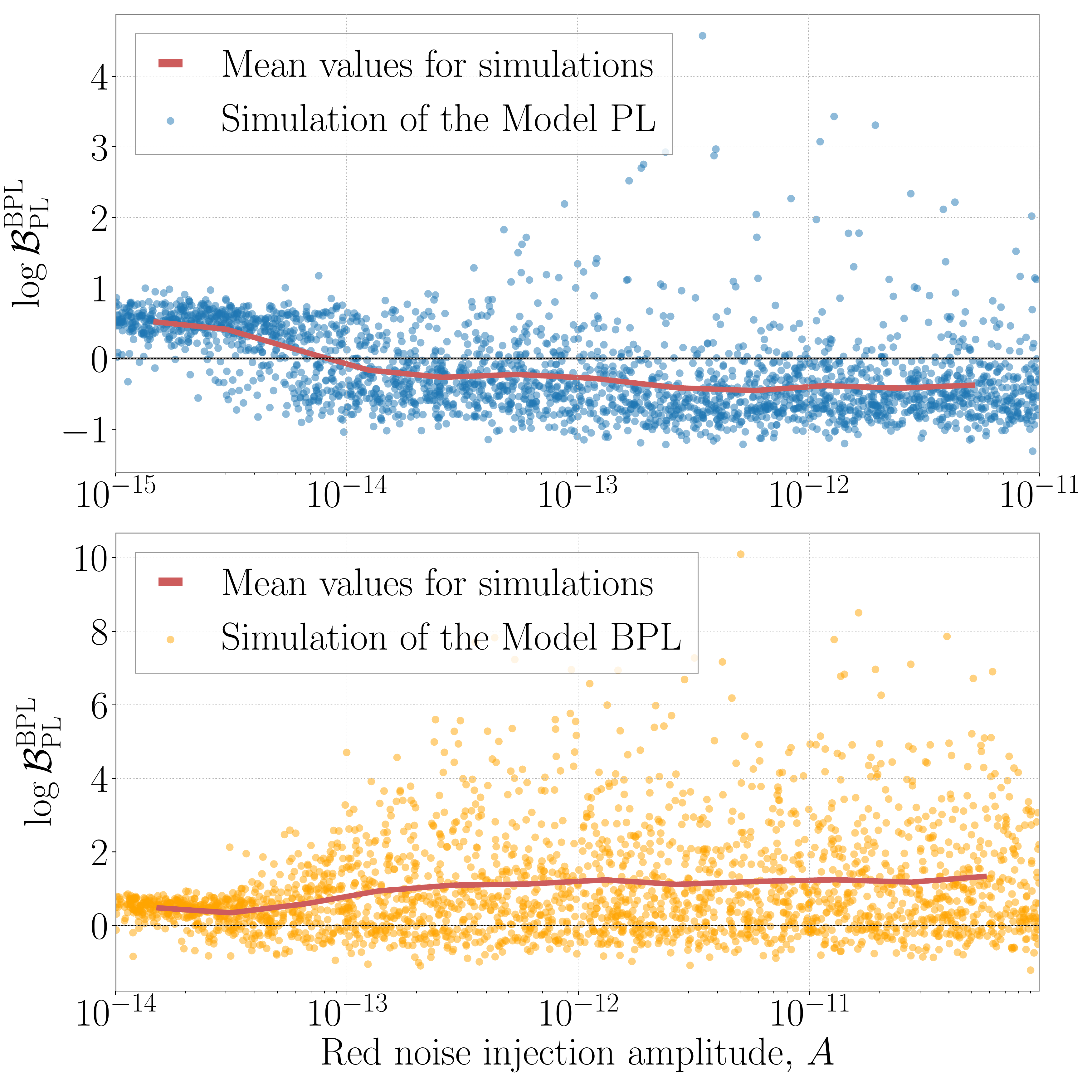}
\caption{
The demonstration of the effect of sample variance on the recovery of a spectral turnover. Each point represents $\log\mathcal{B}^{\text{BPL}}_{\text{PL},i}$. The top plot with blue points is for different realisations of a power law, Model PL (Equation~\ref{eq:default}), while the bottom plot with orange points is for different realisations of a broken power law, Model BPL (Equation~\ref{eq:fancy}). The injection parameters, except red noise injection amplitude $A$ (horizontal axes), are the same for both plots. As the amplitude of the red noise is increased, the evidence in favour (bottom plot) and against (top plot) the spectral turnover plateaus. Red lines are mean values for every 200 simulations.}
\label{fig:sim_variance}
\end{figure}

\begin{table}
\caption{\label{tab:sim_correct}Priors for the injection study in Section~\ref{sec:sim_correct}. Here $\mathcal{U}$ stands for a uniform distribution, and $\text{log}_{10}~\mathcal{U}$ stands for a uniform in $\log_{10}$ distribution.}
\begin{tabular}{l c c c c}
\hline
\begin{tabular}[c]{@{}l@{}}Injected\\ model\end{tabular} &
$\pi(A)$ &
$\log\mathcal{B}^{\text{BPL}}_{\text{PL}}$ &
\begin{tabular}[c]{@{}l@{}}Preferred\\ model\end{tabular}  \\ [1ex] \hline
PL &
\begin{tabular}[c]{@{}l@{}}
    $\log_{10}~\mathcal{U}($ $10^{-14},10^{-12})$ 
\end{tabular} &
$-$30.8 &
PL
\\ [2pt] 
PL &
\begin{tabular}[c]{@{}l@{}}
    $\log_{10}~\mathcal{U}($ $10^{-17},10^{-14})$ 
\end{tabular} &
1.0 &
N/A
\\ [2pt] 
BPL &
\begin{tabular}[c]{@{}l@{}}
    $\log_{10}~\mathcal{U}($ $10^{-14},10^{-12})$ 
\end{tabular} &
95.6 &
BPL
\\ \hline
\end{tabular}
\end{table}

\subsection{Prior mismatch in simulations}
\label{sec:sim_white}
Most of the IPTA pulsars from DR1 are dominated by white noise \cite{lentati2016spin}. 
In this subsection, we perform simulations that demonstrate that model selection for red noise in data, dominated by white noise, can lead to the false detection of a spectral turnover, if we do not carefully choose our prior.
We perform simulations of only white noise with EFAC = 1 and EQUAD = ${\SI{0.1}{\micro \second}}$.
We perform model selection between models BPL and PL.
We observe that evidence for the absence of red noise (Model $\varnothing$) is always the strongest, while either Model PL or BPL may be preferred, depending on our prior on $f_{\text{c}}$ parameter.
As we allow our prior on $f_{\text{c}}$ to include only low values less than around $1/T_{\text{obs}}$, we can not distinguish models PL and BPL.
As we allow our prior on $f_{\text{c}}$ to include only frequencies higher than our sampling frequency, we cannot distinguish between models BPL and $\varnothing$, and model selection between PL and BPL prefers BPL.
This is not surprising, as white noise and Model PL are limiting cases of Model BPL. 
Therefore, for the case of the DR1 analysis, when the true distribution of spin noise parameters is unknown, we propose to account for this effect by including in Equation~\ref{eq:logbftot} only pulsars having $\text{log}~\mathcal{B}_{\varnothing,i}^{\text{PL}} \geq 5$ or $\text{log}~\mathcal{B}_{\varnothing,i}^{\text{BPL}} \geq 5$.
This way we exclude pulsars with no evidence of any spin noise and do not obtain false positives in favor of either a spectral turnover or its absence.
Another solution to this problem is to fit the priors using the hierarchical inference \citep{mackay2003information}, which we defer to a future work.

\subsection{The effect of sample variance in recovery of high amplitude red noise}
\label{sec:sim_red}
In this subsection we find that with a PTA observation time of 10 years, we are unlikely to resolve a turnover in the red noise process of any particular pulsar, assuming a fiducial $f_{\text{c}} = 10$ nHz.
The is because factors $a_i$ and $b_i$ in Equation~\ref{eq:fourierbasis} become a source of noise themselves, and we do not have a data span long enough to effectively probe residuals spectra at frequencies around the turnover.

To demonstrate this, we simulate 1000 pulsars with red noise Model PL amplitude $\pi(A) = \log_{10}\mathcal{U}(10^{-15};10^{-11})$ and $\gamma = 3$, and simulate additional 1000 pulsars with red noise Model BPL with the same parameters and a corner frequency $f_{\text{c}} = 10$ nHz.
As the amplitude of the red noise in the set of simulated pulsars increases, the average $\text{log}\mathcal{B}_i$ in favor of the correct model plateaus. This is demonstrated in Figure~\ref{fig:sim_variance}.
We can see that, at some point, increasing $\log\mathcal{B}(f)$ starts slightly favouring the correct model, but then saturates, so that increasing the amplitude of the red noise does not help to resolve a low-frequency turnover. In this medium-to-strong red noise regime, some realisations of Model PL may favour the Model BPL hypothesis, and vice versa. However, the mean $\log\mathcal{B}^{\text{BPL}}_{\text{PL},i}$ (red line in Figure~\ref{fig:sim_variance}) favours the correct model.

\section{Sources of noise in the first IPTA data release}
\label{sec:noiseipta}
In this Section, we describe sources of noise in the IPTA DR1 dataset.
We use~\cite{lentati2016spin} as a guide for choosing what noise terms to include in our model.
In Table~\ref{tab:priors}, we list the prior distributions for parameters used in our models.
Then we perform Bayesian inference of these parameters and model selection for millisecond pulsar spin noise.

\begin{table}
\caption{\label{tab:priors}Priors used for model selection analyses between models PL (Equation~\ref{eq:default}) and BPL (Equation~\ref{eq:fancy}), and between models PL and M (Equation~\ref{eq:melatos}). Column 2 indicates whether the prior has been used in all model comparison analyses, or in model comparison between specific models.}
\begin{tabular}{l c c}
\hline
Parameter $\bm{\theta}$ & Model comparison & Prior $\pi(\bm{\theta})$ \\ [1ex] \hline
EFAC & all & $\mathcal{U}(0,10)$ \\ [2pt] 
EQUAD [s] & all & $\text{log}_{10}~\mathcal{U}(10^{-10},10^{-4})$ \\ [2pt] 
ECORR [s] & all & $\text{log}_{10}~\mathcal{U}(10^{-10},10^{-4})$ \\ [2pt] 
$A_{\text{SN}}$ & PL-BPL & $\text{log}_{10}~\mathcal{U}(10^{-20},10^{-8})$ \\ [2pt] 
~ & PL-M & $\text{log}_{10}~\mathcal{U}(10^{-17},10^{-10})$ \\ [2pt] 
$\gamma_{\text{SN}}$ & all & $\mathcal{U}(0,10)$ \\ [2pt] 
$f_{\text{c}}$ [Hz] & PL-BPL & $\text{log}_{10}~\mathcal{U}(10^{-12},10^{-6})$ \\ [2pt] 
$M_{\text{SN}}$ & PL-M & $\text{log}_{10}~\mathcal{U}(10^{-1},10^{6})$ \\ [2pt] 
$t_{\text{c}}$ [s] & PL-M & $\text{log}_{10}~\mathcal{U}(2 \pi \times 10^8, 10^{22})$ \\ [2pt] 
$A_{\text{DM}}$ & all & $\text{log}_{10}~\mathcal{U}(10^{-20},10^{-8})$ \\ [2pt] 
$\gamma_{\text{DM}}$ & all & $\mathcal{U}(0,10)$ \\ [2pt] 
$A_{\text{BS}}$ & all & $\text{log}_{10}~\mathcal{U}(10^{-16},10^{-10})$ \\ [2pt] 
$\gamma_{\text{BS}}$ & all & $\mathcal{U}(0,10)$ \\ [2pt] 
$A_{\text{E}}$ & all & $\text{log}_{10}~\mathcal{U}(10^{-10},10^{-2})$ \\ [2pt] 
$t_{\text{E}}$ [MJD] & all & $\mathcal{U}(54500,54900)$ \\ [2pt] 
$\tau_{\text{E}}$ [MJD] & all & $\text{log}_{10}~\mathcal{U}(5,100)$ \\ [2pt] 
$A_{\text{G}}$ & all & $\text{log}_{10}~\mathcal{U}(10^{-6},10^{-1})$ \\ [2pt] 
$t_{\text{G}}$ [MJD] & all & $\mathcal{U}(53710,54070)$ \\ [2pt] 
$\sigma_G$ [MJD] & all & $\mathcal{U}(20,140)$ \\ \hline
\end{tabular}
\end{table}

\subsection{White noise}
\label{sec:whitenoise}
IPTA pulsars are often monitored by several radio observatories. The raw voltages from each telescope are processed by different hardware.
Each observing system has different measurement errors, contributing to measured white noise.
Noise parameter EFAC, introduced in Equation~\ref{eq:efacequad}, accounts for ToA uncertainty, associated with errors during the process of cross-correlation of pulse profile templates with observed pulse profiles.
Parameter EQUAD is introduced to account for stochastic variations in both phase and amplitude of radio pulse profiles.
These variations are called ``pulse jitter''~\citep{oslowski2011jitter,shannon2014jitter}.
Parameters EFAC and EQUAD are introduced for each backend system that processes raw telescope data, in accordance with Equation~\ref{eq:efacequad}.
In NANOGrav data, one epoch of observations with wide-band receivers is split into multiple ToAs, corresponding to different radio-frequencies, or sub-bands.
Thus, for NANOGrav data, ECORR parameters are introduced to account for correlations between sub-banded ToAs at each epoch \citep{11yrlimits2018nanograv}.

\subsection{DM noise}
\label{sec:dmnoise}
Dispersion measure (DM) is the electron column density, integrated along the line of sight to a pulsar. Stochastic variations in dispersion measure result in DM noise.
We model DM noise as a power law with $A_{\text{DM}}$ and $\gamma_{\text{DM}}$, where $\kappa_j = K^2 \nu_j^{-2}$ in Equation~\ref{eq:fourierbasis}.
So, both $\kappa_j$ and $F_{i,j}$ depend on the radio frequency $\nu_j$ (Hz) of the $j$'th ToA.
A constant $K = 1400$ MHz can be thought of as a reference radio frequency.
We account for DM variations for every pulsar in IPTA analysis.

\subsection{Band noise and system noise}
\label{sec:bandsys}
\cite{lentati2016spin} found that specific IPTA pulsars show evidence of band noise and system noise, which introduces additional red noise in some observing systems and radio frequency bands. In order to separate band noise and system noise from spin noise, we add a separate power law with $A_{\text{BS}}$ and $\gamma_{\text{BS}}$ on specific radio frequency bands and systems for specific pulsars where band and system noise for IPTA data release 1 has been found \citep[see Table 4 in][for details]{lentati2016spin}. 

\subsection{Spin noise}
\label{sec:spinnoise}
We model spin noise as a common red noise process between all observing systems and radio frequencies. 
Model PL depends on parameters $A_{\text{SN}}$ and $\gamma_{\text{SN}}$, Model BPL depends on an additional parameter $f_{\text{c}}$.
We refer to a hypothesis that no spin noise is present in the data, as to Model $\varnothing$.
In this work, we are mostly interested in resolving a spectral turnover in spin noise, characterized by the parameter $f_{\text{c}}$ in Model BPL.
We are also interested in Model M with parameters $M_{\text{SN}}$ and $t_{\text{c}}$\footnote{Although Bayes factors can be applied to non-nested models M and PL~\citep{kass1995bayes}, some recent works pointed out difficulties in that approach and possible solutions~\citep{hong2005nonnested}.}.
When carrying out model selection between Model M and Model PL, we chose our prior on Model PL amplitude $A$ to match the range of spin noise amplitudes that is allowed by our priors for $\eta(R^{-1})$ and $\lambda$ in Model M.
Otherwise, the model with a wider prior range on spin noise amplitude would be incorrectly penalized when calculating a Bayes factor.

\subsection{Transient noise events}
\label{sec:transient}
Pulsars J1713+0747 and J1603$-$7202 show evidence of a sudden change in dispersion measure~\citep{coles2015ismscattering, keith2012dispersion, desvignes2016eptatiming, zhu2015testinggr}. We take these events into account using the same empirical models that were used in~\cite{lentati2016spin}. For J1713+0747 we model the event as a frequency-dependent sudden decrease followed by an exponential increase in timing residuals:

\begin{equation}\label{eq:model_expdip}
    s_{\text{E}}( t | A_{\text{E}}, t_{\text{E}}, \tau_{\text{E}}) = K^2 \nu^{-2}
        \begin{cases}
            0 , ~t < t_{\text{E}}~; \\
            A_{\text{E}}~e^{- \frac{t-t_{\text{E}}}{\tau_{\text{E}}} }, ~t \geq t_{\text{E}}~;
        \end{cases}
\end{equation}
where $\nu$ is a radio frequency, and $K = 1400$ MHz is the same reference frequency as we use to model DM noise.
We model the DM event in pulsar J1603$-$7202 as a Gaussian function in the time domain:
\begin{equation}\label{eq:model_gaussian}
    s_{\text{G}}( t | A_{\text{G}}, t_{\text{G}}, \sigma_{\text{G}}) = K^2 \nu^{-2} A_{\text{G}}~e^{- \frac{(t-t_{\text{G}})^2}{2 \sigma_{\text{G}}^2} }.
\end{equation}
DM event models in Equation~\ref{eq:model_expdip} and Equation~\ref{eq:model_gaussian} are added to the signal vector $\bm{s}$ in the likelihood.

\begin{table*}
\caption{\label{tab:results_red}Results for IPTA DR1 pulsars where we found $\text{log}~\mathcal{B}_{\varnothing,i}^{\text{BPL}}>0$ and $\text{log}~\mathcal{B}_{\varnothing,i}^{\text{PL}}>0$.
Columns 3 ($A_{\text{SN}}$) and 4 ($\gamma_{\text{SN}}$) are the red noise parameter estimates for Model PL.
Columns 5 ($\log{\mathcal{B}^{\text{PL}}_{\varnothing,i}}$) and 6 ($\log{\mathcal{B}^{\text{BPL}}_{\varnothing,i}}$) show whether pulsar data favours Models BPL (Equation~\ref{eq:fancy}) and PL (Equation~\ref{eq:default}) against no spin noise. Columns 7 ($\log{\mathcal{B}^{\text{BPL}}_{\text{PL},i}}$) and 8 ($\log{\mathcal{B}^{\text{M}}_{\text{PL},i}}$) show how specific pulsars favors Models BPL and M (Equation~\ref{eq:melatos}) over Model PL. Here we assume a Solar System ephemeris model DE421, which is a default option for IPTA DR1.}
\begin{tabular}{l c c c c c c c }
\hline
Pulsar & $T_{\text{obs}}$ (\text{yr}) & $\text{log}_{10}A_{\text{SN}}$ & $\gamma_{\text{SN}}$ & $\text{log}\mathcal{B}^{\text{PL}}_{\varnothing,i}$ & $\text{log}\mathcal{B}^{\text{BPL}}_{\varnothing,i}$ & $\text{log}\mathcal{B}^{\text{BPL}}_{\text{PL},i}$ & $\text{log}\mathcal{B}^{\text{M}}_{\text{PL},i}$ \\ [1ex] \hline
J0613$-$0200 & 13.7 & $-14.62_{-1.20}^{+0.60}$ & $4.70_{-0.92}^{+2.88}$ & 10.7 & 10.2 & $-0.5$ & $-2.0$ \\ [2pt] 
J0621+1002 & 14.3 & $-12.10_{-0.13}^{+0.12}$ & $2.50_{-0.43}^{+0.72}$ & 4.6 & 6.5 & 1.9 & 1.5 \\ [2pt] 
J1713+0747 & 21.2 & $-14.81_{-0.83}^{+0.39}$ & $4.55_{-0.69}^{+1.90}$ & $>$11.7 & $>$11.6 & $-$0.2 & $-4.8$ \\ [2pt] 
J1824$-$2452A & 5.8 & $-12.80_{-3.05}^{+0.56}$ & $2.30_{-0.32}^{+4.44}$ & 19.0 & 18.8 & $-$0.2 & 1.3 \\ [2pt] 
J1939+2134 & 27.1 & $-14.33_{-0.40}^{+0.24}$ & $6.31_{-0.54}^{+0.80}$ & $>$12.5 & $>$11.4 & $-$1.1 & $-109.8$ \\ [2pt] 
J2145-0750 & 17.5 & $-13.03_{-0.06}^{+0.09}$ & $0.44_{-0.14}^{+0.57}$ & $>$11.6 & $>$12.5 & 0.8 & $-2.0$ \\ [2pt]
J1024$-$0719$~^*$ & 15.9 & $-13.94_{-0.41}^{+0.22}$ & $5.41_{-0.53}^{+1.00}$ & $>$12.4 & $>$11.8 & $-$0.6 & $-29.0$ \\ [2pt] 
\hline
\end{tabular}
\end{table*}


\section{Results}
\label{sec:results}
We perform parameter estimation and model selection for pulsars from the first IPTA data release.
A summary of our analysis for individual pulsars is given in Table~\ref{tab:results_red}.
The first column contains pulsar names and the second column contains observation spans.
The next two columns, $\text{log}_{10}A_{\text{SN}}$ and $\gamma_{\text{SN}}$, represent parameter estimates for Model PL with errors, based on 16\% and 84\% levels of marginalized posteriors.
The last two columns contain the results of spin noise model selection.
From the seventh column, we see that specific pulsars do not show support in favour of a spectral turnover because $| \log \mathcal{B}^{\text{BPL}}_{\text{PL},i} | < 2$ for all pulsars.

Next, we employ Equation~\ref{eq:logbftot}, in order to use all available data for model selection.
We perform our analysis with five different Solar System ephemeris models, as it has been found that errors in Solar System ephemerides contribute to pulsar red noise~\citep{caballero2018studying,11yrlimits2018nanograv,guo2019studying}.
We find that data favours neither Model PL, nor Model BPL. This result is summarized in Table~\ref{tab:results_total}.

Note, Tables~\ref{tab:results_red} and~\ref{tab:results_total} contain only results from seven pulsars where $\log\mathcal{B}^{\text{PL}}_{\varnothing}>5$ or $\log\mathcal{B}^{\text{BPL}}_{\varnothing}>5$.
In Table 6 in \cite{lentati2016spin}, authors present eight pulsars that show evidence for spin noise in their analysis.
Seven of them can be found in our Table~\ref{tab:results_total}: J0613$-$0200, J0621$+$1002, J1713$+$0747, J1824$-$2452A, J1939$+$2134, J2145$-$0750 and J1024$-$0719.
In the remaining pulsar J1012$+$5307 we did find some evidence of spin noise, $\text{log}~\mathcal{B}_{\varnothing,i}^{\text{PL}} = 4.3$, assuming the default Solar System ephemeris DE421.
However, J1012$+$5307 did not satisfy our formal criteria to be included in Table~\ref{tab:results_total}.
It is worth noting that in \cite{lentati2016spin} pulsar J2145$-$0750 is found to have the most shallow power-law index $\gamma_{\text{SN}} = 0.6 \pm 0.2$.
For the reasons discussed in Section~\ref{sec:red_noise_models}, J1012$+$5307 only showed evidence of spin noise in our analysis after we changed a number of Fourier components $N_\text{F}$ from 30 to 100 for this pulsar.
Pulsar J1024$-$0719 is marked with a star for the following reason.
It has been suggested that the spin noise in J1024$-$0719 originates from a companion star in a long-period binary system \citep{kaplan2016longorbitj1024}.
After we take binary motion into account, by adding a second spin frequency derivative into the timing model, we see no evidence for spin noise in J1024$-$0719.

The last column in Table~\ref{tab:results_red}, $\log{\mathcal{B}^{\text{M}}_{\text{PL},i}}$, presents log Bayes factors in favour of Model M over Model PL.
We find that no pulsars show a strong support for Model M. However, pulsars J1939+2134, J1024$-$0719 and J1713+0747 disfavour Model M with $\log{\mathcal{B}^{\text{M}}_{\text{PL},i}} < -4$.

\begin{figure}
\centering
\includegraphics[width=1.0\linewidth]{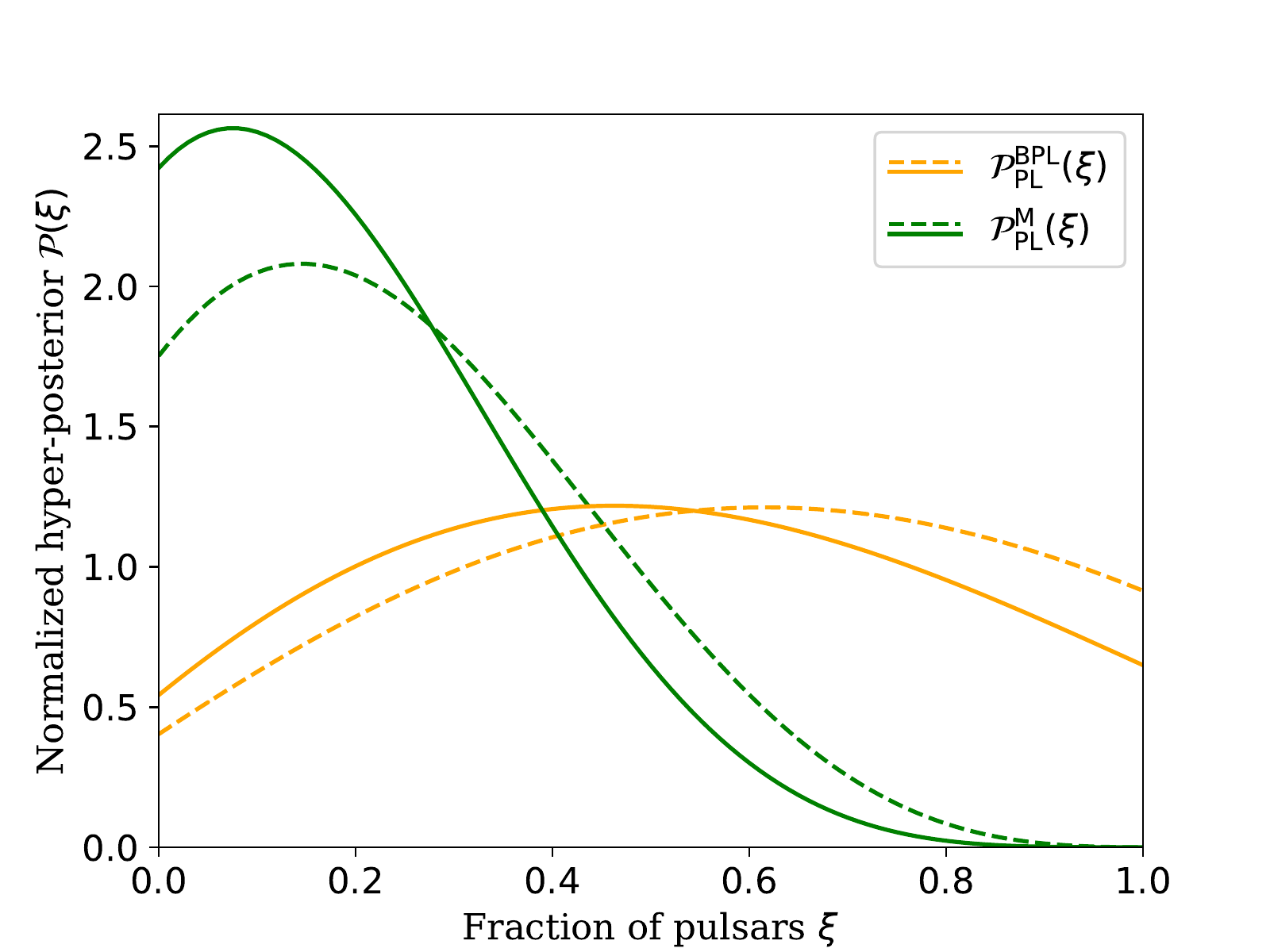}
\caption{
Hyper-posteriors $\mathcal{P}(\xi)$ for DR1 pulsars. Orange lines are posteriors $\mathcal{P}^{\text{BPL}}_{\text{PL}}(\xi)$ for the fraction of pulsars that are described by Model BPL (Equation~\ref{eq:fancy}), assuming other pulsars are described by Model PL (Equation~\ref{eq:default}).
Green lines are posteriors $\mathcal{P}^{\text{M}}_{\text{PL}}(\xi)$ for a fraction of pulsars that are described by Model M (Equation~\ref{eq:melatos}), assuming other pulsars are described by Model PL.
For solid lines, we assume that spin noise in J1024$-$0719 is intrinsic to the pulsar. For dashed lines, we assume that the apparent spin noise in J1024$-$0719 is caused by the second spin frequency derivative of the pulsar induced by gravitational interaction of J1024$-$0719 with a binary companion star~\citep{kaplan2016longorbitj1024}.}
\label{fig:hyper_mixpop}
\end{figure}

We also consider that our data may contain a mixture of pulsars from two models. For this case, we define a likelihood:
\begin{equation}\label{eq:hyperlikelihood}
\mathcal{L}_{\text{B}}^{\text{A}}(\xi) = \prod_{i=1}^{\text{N}_{\text{psr}}}
\bigg(\xi \mathcal{Z}_i^{\text{A}} + (1-\xi) \mathcal{Z}_i^{\text{B}} \bigg)~,
\end{equation}
where $\xi$ is a hyper-parameter that determines the fraction of pulsars that are described by model A. The rest of the pulsars are described by model B.
Using Equation~\ref{eq:hyperlikelihood}, we estimate the fraction of pulsars that are consistent with a superfluid turbulence origin and a spectral turnover. The results are summarized in Figure~\ref{fig:hyper_mixpop}.
We estimate that the fraction of pulsars with the spectral turnover is consistent with any number between 0 and 1, while the fraction of pulsars where Model M is favored over Model PL is mostly consistent with zero.
Since no spectral turnover is detected, pulsars J0621$+$1002 and J1824$-$2452A could get positive preference for Model M over Model PL because their power-law index $\gamma$ is consistent with 2.

\begin{table}
\caption{\label{tab:results_total}The overall $\log\mathcal{B}^{\text{BPL}}_{\text{PL}}$ in favour of Model BPL (Equation~\ref{eq:fancy}) over Model PL (Equation~\ref{eq:default}), using all available IPTA data, for different Solar System ephemeris models.}
\begin{tabular}{l c c}
\hline
Ephemeris & $\log\mathcal{B}^{\text{BPL}}_{\text{PL}}$ & $\log\mathcal{B}^{\text{BPL}}_{\text{PL}}$ (without J1024$-$0719)  \\ [1ex] \hline
%
DE405 & $-0.4$ & $0.3$ \\ [2pt] 
DE418 & $-1.0$ & $-0.3$ \\ [2pt] 
DE421 & $0.2$ & $0.8$ \\ [2pt] 
DE430 & $-0.1$ & $0.7$ \\ [2pt] 
DE435 & $-0.8$ & $-0.1$ \\ 
\hline
\end{tabular}
\end{table}

\section{Conclusions}
\label{sec:conclusion}
We perform Bayesian model selection to search for a spectral turnover in pulsar spin noise using the first data release of the IPTA.
We find support, with a log Bayes factor above 4, for spin noise in eight pulsars, which is consistent with \citet{lentati2016spin}.
However, we find no evidence for a spectral turnover either in individual pulsar data or by combining different pulsars.
We also fit the data to the superfluid turbulence model proposed by \citet{melatos2013superfluid}.
Our results show that whereas this model is indistinguishable from the power-law model for most pulsars, it is strongly disfavored by three pulsars, especially PSR J1939+2134 with a log Bayes factor of 110.

Based on a range of simulations, we find that one is unlikely to resolve a spectral turnover with a fiducial corner frequency of 10 nHz in any pulsar with $\approx 10$ years of observations.
Longer data spans are required to increase the detection confidence of a spectral turnover in individual pulsars, while a larger number of pulsars with red noise can help to resolve the presence of a spectral turnover in a population of pulsars.
A follow-up study using longer data sets and a larger sample of pulsars, e.g., the IPTA second data release \citep{IPTAdr2}, will prove useful in not only understanding the nature of red noise in millisecond pulsars but also in evaluating the realistic prospect of gravitational-wave detection.
A more detailed simulation study is required to explore pulsar timing array configurations that would resolve spectral turnover in the individual pulsars. Whereas our simulation study assumed a pulsar with observation span of 10 years, two pulsars from the first data release of the IPTA have observations spans above 25 years. At the same time, next-generation pulsar timing arrays based on MeerKat, FAST, SKA, will have a reduced radiometer noise. Both greater observation spans and reduced white noise levels will increase the sensitivity of a pulsar timing array to the spectral turnover, and the future study could help to estimate by how much. Simulations that attempt to provide a precise answer to these questions for the real data ought to include all other noise sources (i.e., DM noise, band noise), multiple observing backends with realistic observation cadences.
Another interesting future simulation study would determine whether the broken power-law model would be favored over the power-law model when the superfluid turbulence model is simulated.

\section{Acknowledgements}
\label{sec:acknowledgements}
We thank the anonymous referee for valuable feedback.
We thank Nataliya Porayko, Yuri Levin, Daniel Reardon and Paul Lasky for useful discussions.
BG, XZ, and ET are supported by ARC CE170100004.
ET is additionally supported by ARC  FT150100281.

\section*{Data availability}
\label{sec:dataavailability}
The first data release of the IPTA underlying this article is available at \href{https://gitlab.com/IPTA/DR1}{gitlab.com/IPTA/DR1}.
The simulated data underlying this article will be shared on a reasonable request to the corresponding author.





\bibliographystyle{mnras}
\bibliography{mybib} 





\appendix
\section{Explicit form of Model M power spectral density}
\label{sec:appendix}
The definite integral in Equation~\ref{eq:melatos} yields an analytical solution. First, we reparametrize Equation~\ref{eq:melatos}:
\begin{equation}\label{eq:reparam_melatos}
\begin{gathered}
\begin{cases}
M = \frac{15}{(4 \pi \lambda)^2}~; \\
t_{\text{c}} = \frac{2 \pi}{\eta(R^{-1})}~.
\end{cases}
\end{gathered}
\end{equation}
Next, we obtain the analytical solution in a form:
\begin{equation}\label{eq:analytical_melatos}
\begin{gathered}
    P(f) = \frac{3 M p^2}{4 t_{\text{c}} f^2}
    \Bigg(
        \frac{1}{128 \sqrt[3]{2} \pi^{16/3}} + 
        \frac{3}{704 \sqrt[3]{2} \pi^{22/3}} + 
        \frac{9}{3584 \sqrt[3]{2} \pi^{28/3}}
        - \\
        \frac{1}{f^2 t_{\text{c}}^2} \bigg(
            \frac{1}{48 \pi^4} + 
            \frac{1}{96 \pi^6} + 
            \frac{3}{512 \pi^8}
        \bigg)
        + \\
        \frac{1}{f^4 t_{\text{c}}^4} \bigg(
            \frac{1}{8 2^{2/3} \pi^{8/3}} + 
            \frac{3}{56 2^{2/3} \pi^{14/3}} +
            \frac{9}{320 2^{2/3} \pi^{20/3}}
        \bigg)
        - \\
        \frac{1}{f^6 t_{\text{c}}^6} \bigg(
            \frac{1}{2 \sqrt[3]{2} \pi^{4/3}} + 
            \frac{3}{20 \sqrt[3]{2} \pi^{10/3}} + 
            \frac{9}{128 \sqrt[3]{2} \pi^{16/3}}
        \bigg)
        + \\
        \frac{1}{f^8 t_{\text{c}}^8} \bigg(
            \frac{1}{2 \pi^2} + 
            \frac{3}{16 \pi^4} -
            \frac{4 \log(\pi)}{3} -
            \log{4}
        \bigg)
        - \\
        \frac{1}{f^{10} t_{\text{c}}^{10}} \bigg(
            \frac{3 \sqrt[3]{2}}{\pi^{2/3}}+
            \frac{9}{8 2^{2/3} \pi^{8/3}}
        \bigg)
        + \\
        \frac{1}{f^{11} t_{\text{c}}^{11}}
        6 \tan^{-1} \bigg(
            \frac{t_{\text{c}} f}{(2 \pi)^{2/3}}
        \bigg)
        +
        \frac{1}{f^{12} t_{\text{c}}^{12}}
        \frac{9}{2 \sqrt[3]{2} \pi^{4/3}}
        + \\
        \frac{1}{f^{14} t_{\text{c}}^{14}}
        \bigg(
            12 \log(\pi) + 3 \log (64)
        \bigg)
    \Bigg)
\end{gathered}
\end{equation}

\section{Posterior probability distribution examples}
\label{sec:appendix_posteriors}
\begin{figure*}
    \centering
    \begin{subfigure}[b]{0.4\textwidth}
        \includegraphics[width=\textwidth]{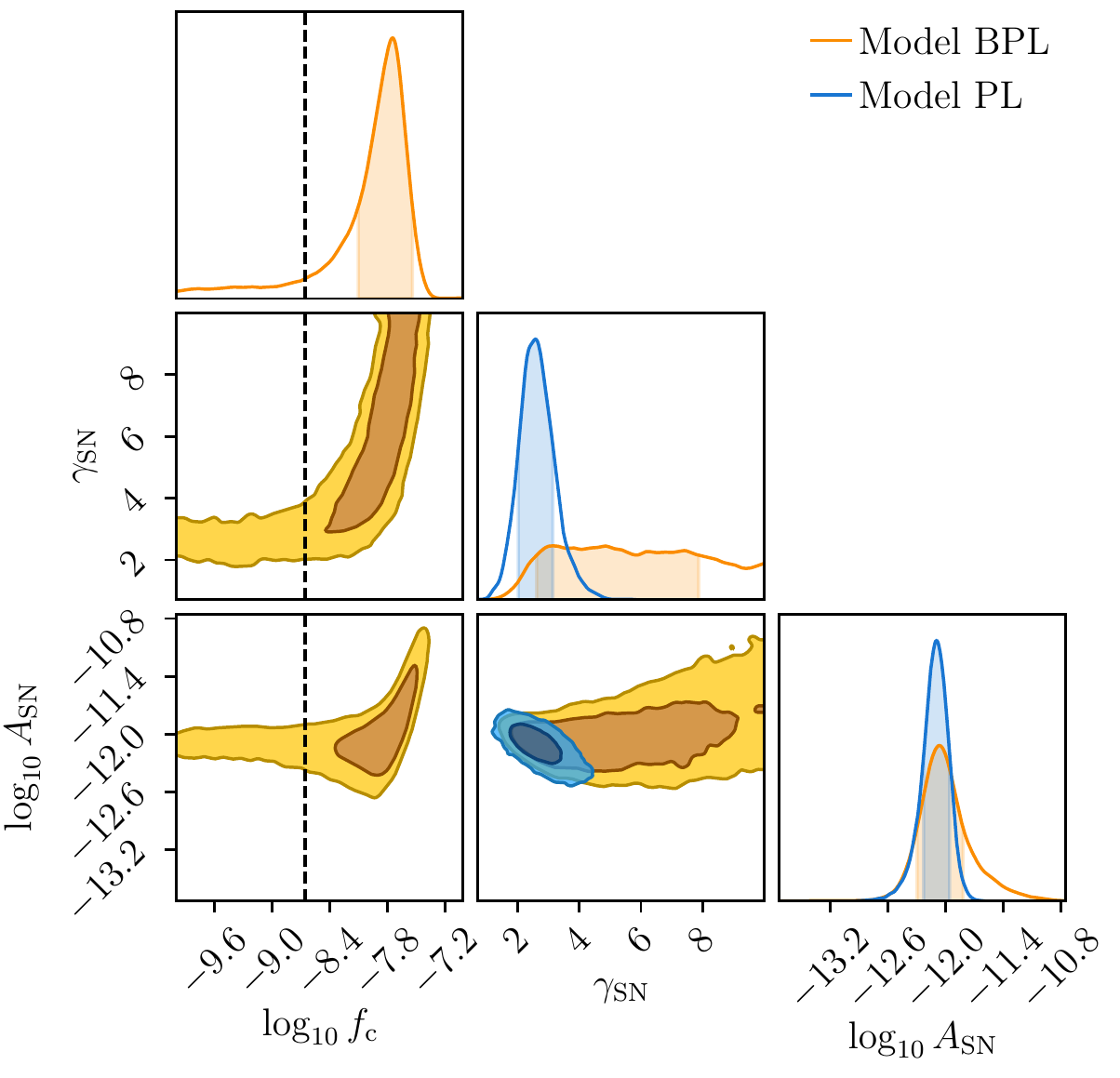}
       \caption{J0621$+$1002}
        \label{fig:0621}
    \end{subfigure}
    ~ 
    \begin{subfigure}[b]{0.4\textwidth}
        \includegraphics[width=\textwidth]{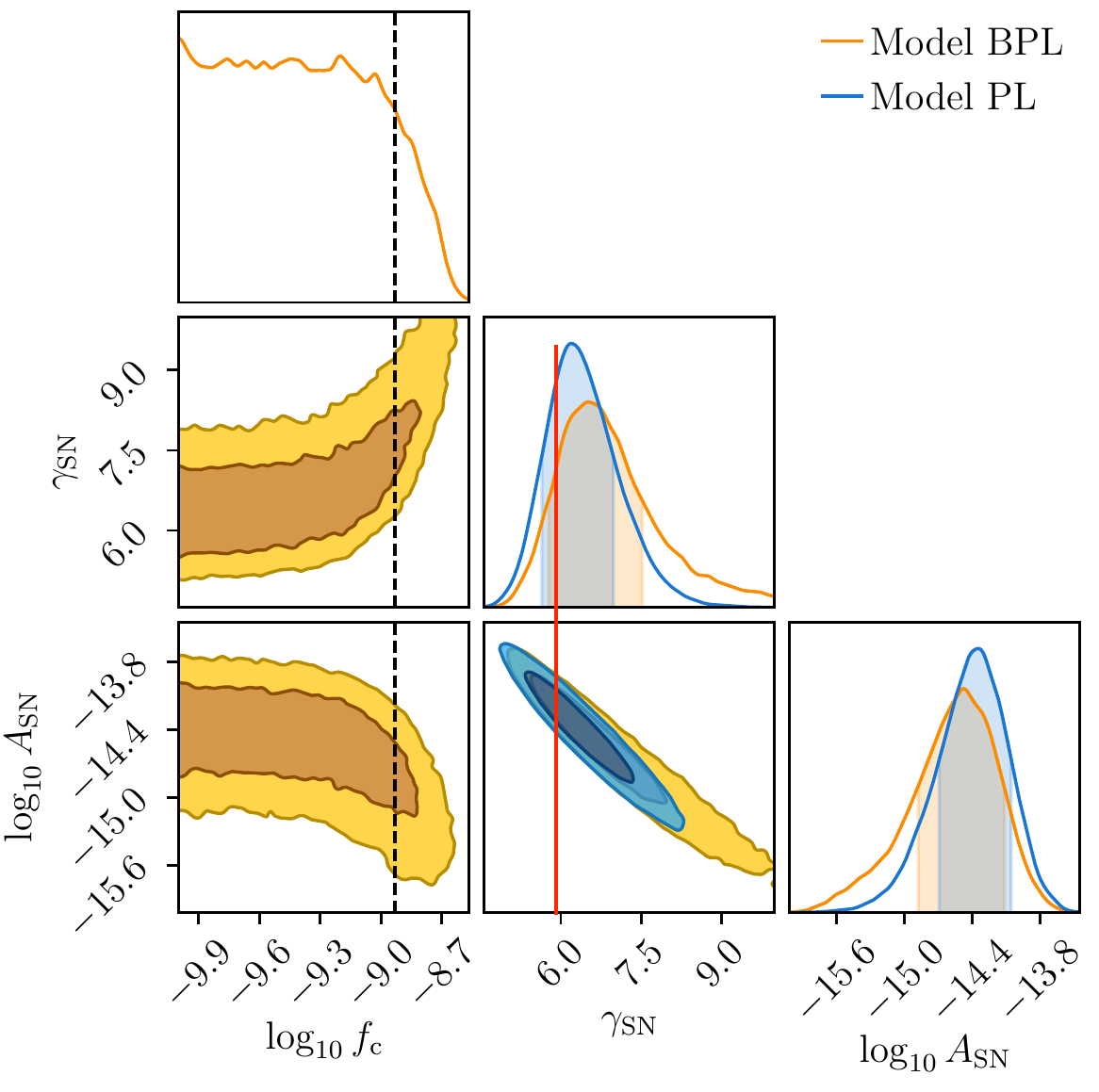}
        \caption{J1939$+$2134}
        \label{fig:1939}
    \end{subfigure}
    ~ 
    \caption{Figure~\ref{fig:posteriors} represents posterior distributions for spin noise parameters for J0621$+$1002 (left, \ref{fig:0621}) and J1939$+$2134 (right, \ref{fig:1939}). Vertical dashed lines represent $1/T_{\text{obs}}$. For J1939$+$2134, with least evidence for the spectral turnover in Table~\ref{tab:results_red}, measurement of $f_{\text{c}}$ does not affect the measurement of the amplitude and the slope of spin noise. However, for J0621$+$1002, with the highest evidence for the spectral turnover in Table~\ref{tab:results_red}, measurement of $f_{\text{c}}$ does affect measurement of the power-law index.}
    \label{fig:posteriors}
\end{figure*}

In Figure~\ref{fig:posteriors} we demonstrate posterior distributions for spin noise parameters of two pulsars, J0621$+$1002 and J1939$+$2134, where the highest and the lowest $\log \mathcal{B}_{\text{PL}}^{\text{BPL},i}$ are found (see Table~\ref{tab:results_red} for details).

\bsp	
\label{lastpage}
\end{document}